\documentclass[%
 reprint,
 amsmath,amssymb,
 aps,
]{revtex4-2}

\usepackage{graphicx}
\usepackage{dcolumn}
\usepackage{bm}

\begin{document}

\preprint{APS/123-QED}

\title{Landau Damping in the Transverse Modulational Dynamics \\ of Co-Propagating Light and Matter Beams}

\author{Christopher Limbach}
\email{Email: climbach@tamu.edu}
\affiliation{%
 Texas A\&M University \\
 College Station, TX 77843 \\
}%




\date{\today}

\begin{abstract}
The optomechanical coupling and transverse stability of a co-propagating monochromatic electromagnetic wave and mono-energetic beam of two-level atoms is investigated in the collisionless regime. The coupled dynamics are studied through a Landau stability analysis of the coupled gas-kinetic and paraxial wave equations, including the effect of the electronic nonlinearity. The resulting dispersion relation captures the interaction of kinetic and saturation effects and shows that for blue detuning the combined nonlinear interaction is unstable below a critical wavenumber which reduces to the result of Bespalov and Talanov \cite{Bespalov1966} in the limit of a negligible kinetic nonlinearity. For red detuning we find that under a saturation parameter threshold exists whereby the system stabilizes unconditionally. With negligible saturation, an optomechanical form of Landau damping stabilizes all wavenumbers above a critical wavenumber determined by the combined strength of the kinetic and refractive optomechanical feedback. The damping is mediated primarily by atoms traveling along the primary diagonals of the Talbot carpet.

\end{abstract}

\maketitle


\section{Introduction\label{sec:intro}} 
The mutual coupling of atomic motion and light propagation, mediated by optical forces and spatial variation in the refractive index, has been studied theoretically \cite{Tesio2013,Tesio2014}, and demonstrated experimentally using cold atomic clouds by retro-reflection feedback \cite{Labeyrie2014} and by structured light a cavity with strong viscous damping \cite{Baio2020}. Recently, spontaneous pattern formation was also demonstrated in free-space laser propagation through a stationary atomic lattice \cite{Schmittberger2016a}. Key features of the optomechanical system include the important role of the Talbot length in selecting the most unstable wavenumber and the emergence of spontaneous symmetry breaking and pattern formation \cite{Camara2015,Labeyrie2014}. In these systems, the kinetic feedback and electronic two-level nonlinearity interact through the saturation parameter, which controls the occurrence and growth-rate of instabilities \cite{Tesio2014}. 

This work examines the role of electronic and kinetic phenomena in the dynamics of the continuous optomechanical interaction, where a free monochromatic wave propagates through a co-moving atomic medium. The continuous system has been studied in the context of atom lasers \cite{Khaykovich2002}, which introduces new phenomena of the condensed state at high phase-space density \cite{Saffman1998}. In this regime the coupled equations permit self-focusing and self-trapping of both matter and optical waves \cite{Saffman2001}. However, even when the atomic motion is incoherent and the dynamics may be described by the gas kinetic equation, the dipole response can strongly couple the field and atomic motion. On this basis Askar'yan first analyzed optomechanical solitons and nonlinear optical guiding of high intensity beams in air \cite{Askaryan1962}. 

In many contexts only a one-way coupling exists between the atomic motion and optical field. For instance, in atomic beam focusing \cite{Bjorkholm1978,Bjorkholm1980,Pearson1980} and guiding experiments \cite{Helseth2002,Riis1990,Arlt2000,Song1999} the refractive index of the atomic medium does not significantly perturb the vacuum field. This was shown recently in particle simulations  used to recreate the atom focusing experiment of Ashkin and co-workers \cite{Bjorkholm1978,Bjorkholm1980,Pearson1980}, wherein an co-propagating and focused laser was observed to focus an atomic beam. By self-consistently including the dipole forces and refractive index, Kumar et al reproduced the previous experimental results in sodium and observed self-focusing of both atomic and optical beams only at higher atomic density, though below the density for condensation \cite{Kumar2021a}. This work quantitatively confirmed the potential for laser intensification due to the refractive index first predicted theoretically by Klimontovich \cite{Klimontovich1979}. Here, we are motivated to understand the transverse instabilities that may arise in such co-propagating beams and the possibility for symmetry breaking, and filamentation.

In the following, we take the approach of Tesio \cite{Tesio2014} and aim to determine the stability of the continuous optomechanical system to small perturbations on a homogeneous background. The results of this approach are thought to apply to beams of transverse size much greater than the critical wavelength of the instability. Section \ref{sec:background} prefaces subsequent analysis by introducing the continuous optomechanical system and linearizing about the homogeneous solution. The stability analysis in Section \ref{sec:stability} then leads to the dispersion relation, from which the critical wavenumber is obtained. For a Cauchy (Lorentzian) velocity distribution the dispersion relation can be explicitly evaluated, from which the poles of the system are obtained and discussed. Numerical solution of the velocity-space perturbation in Section \ref{sec:damping} reveals that kinetic damping occurs through the interaction of resonant particles in a mechanism analogous to Landau damping but with the resonant particle velocity matched to diagonals of the Talbot carpet.
\vspace{-10pt}
\section{Background} \label{sec:background}

We begin by writing the equations for the field and atoms. Above the condensation temperature, where the atomic motion is incoherent, the dynamics are described by the Fokker-Planck equation obtained through adiabatic elimination of the ``fast'' internal dynamics (see \cite{Foot2005,Dalibard1985a}). Coupling between the atoms and co-propagating field is mediated by the dipole response and manifest through the dipole (gradient) force and refractive index. In the following, we examine detunings $\delta/\Gamma >> 1$ sufficient to neglect attenuation by scattering and absorption as well as the corresponding recoil and momentum-space diffusion \cite{Tesio2014}.
\vspace{-10pt}
\subsection{Atomic Beam}
In a sufficiently rarefied environment where collisions may be neglected, the atomic beam evolves by the gas-kinetic Boltzmann equation and we seek stationary solutions for a mono-energetic beam propagating along the $z$ axis. The initial value problem can be written as,
\begin{equation}
    v_b \frac{\partial f}{\partial z} + \mathbf{v}\cdot\frac{\partial f}{\partial \mathbf{x}} + \frac{\mathbf{F}_d}{M} \cdot \frac{\partial f}{\partial \mathbf{v}} = 0.
\label{eq:boltz}
\end{equation}
Here the coordinates $\mathbf{x}$, $\mathbf{v}$ and dipole force $\mathbf{F}_d$ refer only to the transverse plane. The homogeneous solution of Eq. \ref{eq:boltz} consists of a uniform atomic density $N_0$ and refractive index $n_0$ and a distribution function that depends only on the velocity magnitude as $f_0(v)$. 
For a system near-resonance, with not too high saturation or perpendicular velocity, adiabatic elimination of the internal state leads to a gradient force given by \cite{Tesio2014},
\begin{equation}
    \mathbf{F}_d = -\nabla U_d = -\frac{1}{2} \hbar \delta (1+s)^{-1} \nabla_{\perp}s
\label{eq:force}
\end{equation}
where $s = I / I_{sat} (1 + 4 (\delta/\Gamma)^2)$ is the saturation parameter. It should be noted that special care is needed when accounting for the constitutive relations of the medium moving with $v_z = v_b$. In the limit $v_b << c$ the refractive index corresponds to that evaluated in the moving reference frame with corrections of order $(n-1) (v_b/c)^2 << 1$ (see \cite{Kong1986}). 
\vspace{-10pt}
\subsection{Laser Beam}
The laser beam is modeled at a freely traveling, monochromatic classical electromagnetic wave. Applying the slowly varying envelope approximation, we take a trial solution $\tilde{E} = E(\mathbf{r})e^{i(k_0 n_0 z - \omega t)}$ for the wave equation, resulting in the paraxial wave equation for a dilute medium ($n - 1 << 1$) \cite{Born2000}:
\begin{equation}
    2ik_0 n_0 \frac{\partial E}{\partial z} + \nabla^2_{\perp} E + 2k_0^2(n-n_0)E = 0
\label{eq:paraxial}
\end{equation}
where $E(\bf{r})$ is the slow complex field amplitude. In the limit of far detuning, where losses are negligible, the refractive index of two-level atoms is described by \cite{Tesio2014},
\begin{equation}
    n = 1 - \frac{3\lambda^3}{4\pi^2} \frac{\delta/\Gamma}{1 + 4 \delta^2/\Gamma^2} \frac{N}{1+s},
\label{eq:refractive}
\end{equation}
where $N$ is the gas number density which couples to the atomic motion in Eq. \ref{eq:boltz} as $N = \int{f d\mathbf{v}}$. The background refractive index $n_0$ is thus obtained under the conditions $N_0$ and $s_0$, which correspond to the background density and saturation parameter, respectively. Note that the trial solution corresponds to the homogeneous solution for a Kerr medium in the case $n_0 = n_0' + n_2 I_0$ where $n_0'$ is the unsaturated background index and we expand $(1+s)^{-1} \approx 1-s$ for $s<<1$ in Eq. \ref{eq:refractive} (see e.g. \cite{Boyd2009}). 
%
\vspace{-10pt}
\section{Stability Analysis} \label{sec:stability}
To determine the stability of the homogeneous state we expand the distribution function as $f = f_0(\mathbf{v}) + f_1(z,\mathbf{x},\mathbf{v})$ and the field as $E = E_0 + E_1(z,\mathbf{x})$. Inserting the expansion of the distribution function into Eq. \ref{eq:boltz} and eliminating the zeroeth order terms yields the linearized equation, 
%
%
%
%
%
%
\begin{equation}
   v_b \frac{\partial f_1}{\partial z} + \mathbf{v} \cdot \frac{\partial f_1}{\partial \mathbf{x}} - \frac{U_0 \gamma}{M s_0} \nabla_{\perp} s_1 \cdot \frac{\partial f_0}{\partial \mathbf{v}} = 0.
\label{eq:boltz_lin}
\end{equation}
where the parameter $\gamma = s_0 / [(1+s_0) $ln$(1+s_0)]$ is obtained from linearization of the dipole potential. From linearization of the saturation parameter we also find $s_1 / s_0 = (E_1^* + E_1) / E_0$ which introduces the coupling of $E$ and $E^*$ noted in \cite{Boyd2009,Tesio2014} \footnote{In contrast to Eq. \ref{eq:paraxial}, the density perturbation does not couple to the gas-kinetics because the macroscopic and microscopic fields are approximately equal in a dilute medium \cite{Milonni2010}, that is $s \propto I \propto n |E|^2 \approx n_0 |E|^2$.}.

Next we treat the field equation by first expanding the refractive index described by Eq. \ref{eq:refractive} as $n = n_0 + n_1$. Using the expansions of $f$ and $s$ we obtain,
\begin{equation}
        n_1 = (n_0-1) \left( \frac{N_1}{N_0} - \frac{s_1}{1+s_0} \right),
\label{eq:reflin}
\end{equation}
where $N_1 = \int{f_1 d\mathbf{v}}$ is the density perturbation. Note that the first term represents the kinetic coupling while the second term corresponds to the electronic nonlinearity. Substituting this result into Eq. \ref{eq:paraxial} yields,
\begin{equation}
\begin{split}
    2ik_0 n_0 \frac{\partial E_1}{\partial z} + \nabla^2_{\perp} E_1 + \\ 2k_0^2 (n_0-1) \left( \frac{N_1}{N_0} - \frac{s_1}{1+s_0} \right) E_0 = 0.
\end{split}
\label{eq:parax_lin}
\end{equation}
%
%
%
%
%
\section{Landau Stability Analysis}
The stability to perturbations is analyzed by taking the Fourier transform in space and Laplace transform in time of Eq. \ref{eq:parax_lin} and Eq \ref{eq:boltz_lin}, where in the usual way $\partial_x \rightarrow i\mathbf{q}$ and $\partial_z \rightarrow s - g(0)$, where $g(0)$ is the initial condition at $z = 0$ and $s$ is the Laplace transformed variable not to be confused with the saturation parameter. The use of the Laplace transform correctly poses the stability analysis as an initial value problem in space, although for a stationary mono-energetic beam space maps directly to time in the atomic frame. The resulting linear equation for the Fourier-Laplace transforms of the perturbations $\hat{f}_1$ and $\hat{E}_1$, is given by,
\begin{equation}
   v_b s \hat{f}_1 +  i\mathbf{q} \cdot \mathbf{v} \hat{f}_1 - i \frac{U_0 \gamma}{M} \frac{\hat{E}_1 + \hat{E_1}^*}{E_0} \mathbf{q} \cdot \frac{\partial f_0}{\partial \mathbf{v}} = v_b \hat{f}_1(0).
\label{eq:FL_boltz}
\end{equation}
The corresponding equation for the transformed field $\hat{E}_1$ is then,
\begin{equation}
\begin{split}
    2 i k s \hat{E}_1 - q^2 \hat{E}_1 + 2k_0^2 (n_0-1) \\ 
    \left[ \frac{E_0}{N_0} \int f_1 d\mathbf{v} - \frac{s_0}{1+s_0} (\hat{E_1} + \hat{E}_1^*) \right]= 2i k \hat{E}_1(0),
    \end{split}
\label{eq:FL_parax}
\end{equation}
where $\hat{E}_1(0)$ is the Fourier transform of the initial field perturbation. A solution for $\hat{f}_1$ in obtained in terms of the unknown field components $\hat{E}_1$ and $\hat{E}_1^*$ from Eq. \ref{eq:boltz_lin} and then substituted into Eq. \ref{eq:FL_parax} leading to,
%
%
%
%
\begin{equation}
\begin{split}
    (2ik s - q^2) \hat{E}_1 - \\ 2k_0^2 (n_0 - 1) 
    \left[\frac{U_0 \gamma }{M} C_1 - \frac{s_0}{1+s_0} \right]  (\hat{E}_1 + \hat{E}_1^*) \\
    = 2ik_0\hat{E}_1(0) - 2k_0^2 (n_0 - 1) E_0 C_2,
\end{split}
\label{eq:bigeq}
\end{equation}
%
where the kinetic integrals are given by,
\begin{equation}
    C_1 = \int{\frac{i \mathbf{q} \cdot \hat{f}_0}{i\mathbf{q}\cdot \mathbf{v} + v_b s} d\mathbf{v}},
\end{equation}
\begin{equation}
    C_2 = \int{\frac{ v_b \hat{f}_1(0)}{i\mathbf{q}\cdot \mathbf{v} + v_b s} d\mathbf{v}}.
\end{equation}
The integrals $C_1$ and $C_2$ represent the interaction between different velocity groups in the initial distribution function $f_0$ and the perturbation $f_1$ in wavespace. In particular, certain velocity groups will experience a resonant interaction which occurs for those velocities which minimize the denominator. 

At this point, it is convenient to identify key dimensional groups and recast the equations in dimensionless form. We define the diffraction length $L_d = 2k_0/q^2$, which is the (primary) Talbot length reduced by $2\pi$ \cite{Cowley1995}, and re-scale the spatial frequency $s$ as $\bar{s} = s L_d$. We also define the field-atom coupling parameters $\beta = (U_0 \gamma / M)C_1 - s_0/(1+s_0)$ and $V^2 = 2k_0^2(n_0-1)/q^2$. The parameter $V$ is defined analogously to the the fiber V-parameter with core radius $2\pi/q$, but differs from convention by a factor of $(2\pi)^{-2}$. 

Since Eq. \ref{eq:bigeq} includes both an unknown amplitude and phase, a second equation is needed to obtain a solution for $\hat{E}_1$. This is found by taking the complex conjugate of Eqs. \ref{eq:boltz_lin} and \ref{eq:parax_lin} before taking the Fourier-Laplace transform. The resulting pair of linear equations are then cast in the form $\mathbf{M} \mathbf{x} = \mathbf{b}$ as,
\begin{equation}
      \begin{bmatrix}
          i\bar{s}-1 + V^2 \beta &  V^2 \beta \\
           V^2 \beta^* & -i\bar{s}^* - 1 + V^2 \beta^* 
      \end{bmatrix}
       \begin{bmatrix}
          \hat{E}_1 \\
          \hat{E}_1^*
       \end{bmatrix} 
       =
       \begin{bmatrix}
          b \\
          b^*
       \end{bmatrix} 
\label{eq:system}
\end{equation}
where the initial perturbation is given by $b = 2ik/q^2 \hat{E}_1(0) - V^2 E_0 C_2$. Recognizing that the instability response will be dominated by the poles of $\hat{E}_1$ and $\hat{E}^*_1$, the dispersion relation is derived from the characteristic equation of the matrix $\mathbf{M}$ in Eq.~\ref{eq:system}. The dispersion relation for the continuous optomechanical system can be simplified to,
\begin{equation}
    \bar{s}^2 + 1 - 2V^2\beta = 0.
\label{eq:dispersion}
\end{equation}
Notably, the coupling parameter $\beta$ depends on $s$ through the dispersion integral $C_1$ and captures both the kinetic and electronic response. In the absence of the optomechanical coupling ($V = 0, \beta = 0$) the roots are simply $\bar{s} = \pm i$ and represent the well-known marginally stable diffractive oscillation at the Talbot length. 
%
%
\vspace{-10pt}
\subsection{Critical Wavenumber}
First we consider the critical wavenumber corresponding to $\bar{s} = 0$. As shown later, the critical wavenumber is also the threshold wavenumber, that is $Re(s)$ first becomes greater than zero (instability growth) also at $Im(s) = 0$. In the critical case we may evaluate the dispersion integral immediately for a Maxwellian velocity distribution as $ - M N_0 / k_b T_{\perp}$, whereby $\beta = - U_0 \gamma/(k_b T_{\perp}) - s_0/(1+s_0)$ and the critical wavenumber $q_c$ is,
\begin{equation}
    q_c^2 = 4 k_0^2(n_0-1) \left( - \frac{U_0}{k_b T_{\perp}} - \frac{s_0}{1+s_0} \right). 
\label{eq:critical}
\end{equation}
Recall that for red detuning we have $U_0 < 0$ and $n_0 - 1 > 0$ while for blue detuning $U_0>0$ and $n_0-1 < 0$, whereas $s_0$ is always positive. On this basis we find that red and blue detuning exhibit qualitatively different behavior. For blue detuning there always exists a critical wavenumber, as both the kinetic and electronic nonlinearities are positive. That is, the repulsion of atoms out of high intensity regions of the beam reinforces the self-focusing Kerr nonlinearity, both serving to increase the refractive index. From this result we can directly recover the critical wavenumber of the Bespalov-Talanov instability by identifying $2 (n_0-1) s_0/(1+s_0) \rightarrow n_2/n_0 |E|^2$ (see \cite{Boyd2009} pp. 207).  

In the case of red-tuning the kinetic term contributes to instability while the saturation effect provides a stabilizing influence and unconditionally stabilizes the system at the threshold,
\begin{equation}
    s_{th} = \frac{ |U_0|/k_b T_{\perp}}{1 - |U_0|/k_b T_{\perp}}.
    \label{eq:s_thresh}
\end{equation}
 Since the saturation term in Eq. \ref{eq:critical} cannot rise above unity, we find that for a strong mechanical interaction ($|U_0| > k_B T$) the system cannot be stabilized. 
 
Further examination of Eq. \ref{eq:critical} shows that the dephasing provided by higher temperature lowers the threshold wavenumber but does not eliminate it. When $s_0 \rightarrow 0$ and $U_0 \sim k_b T_{\perp}$, we can compare the results above with diffractive coupling to a stationary atomic cloud by mirror back-reflection, for which the critical wavenumber is $q_c = \sqrt{\pi k_0/2d}$, where $2d$ is the length of the propagation \cite{Tesio2014}. In this arrangement, the dominant structures were of size $a_c$ with a Fresnel number near unity, that is with $q_c = 2\pi / a_c $, where the Fresnel number is $N_f = a_c^2/(2d \lambda) = 2$. This is analogous to atmospheric optics where the strongest irradiance fluctuations at some propagation length $L$ correspond to atmospheric eddies for which $N_f \sim 1$ \cite{Tyson2000}. In the co-propagating beams examined here, the diffractive length scale is rather set by the refractive index, where the relevant diffractive length is $(k_0|n_0 - 1|)^{-1}$. By analogy to fiber optics these are structures with $V^2 \sim 1$ based on their size, whereas larger structures $V^2 >> 1$ are unstable and spontaneously break into filamentary structures \cite{Saleh1991,Boyd2013,Boyd2009}. Thus, we find that the threshold wavenumber is dependent on the combination (product) of the refractive feedback $n_0 - 1$ and kinetic feedback, either through the gradient force response $U_0/k_b T_{\perp}$ or the electronic nonlinearity.
\vspace{-10pt}
\subsection{Growth Rate and Damping}
The determination of growth rates proceeds from solving the dispersion relation Eq.~\ref{eq:dispersion}. However, the kinetic integral $C_1$ cannot be evaluated directly for a Maxwellian velocity distribution, although \cite{Tesio2014} showed that instability growth rates for a Maxwellian and Cauchy (Lorentizan) velocity distribution were nearly identical, provided the characteristic width $v_{th}$ of the distributions were the same. 
%
%
%
%

We therefore proceed by analyzing a gas with an isotropic Cauchy velocity profile which permits an analytical solution of $C_1$. Following \cite{Gurnett2017}, the integral along the real axis in the $v$-plane can completed by contour integration resulting in,
%
\begin{equation}
    C_1 = -\frac{1}{v_{th}^2} (1 + \bar{s}\tau)^{-2},
\label{eq:c1_lorentz}
\end{equation}
where we find a new dimensionless parameter $\tau = v_b |q| / 2 k_0 v_{th}$. This key parameter represents the ratio between the diffraction angle $\theta_d \sim q/k_0$ and the thermal spreading angle $v_{th}/v_b$. It can also be interpreted as a ratio of characteristic distances $z_b/z_T$, where $z_b = v_b / v_{th}|q|$ is the distance over which a particle with velocity $v_{th}$ moves a transverse distance $2\pi/|q|$ and $z_T$ is the Talbot length.

The dispersion relation can then be recast as a function of only two dimensionless parameters by noting that the product $2 V^2 \beta$ can be rewritten as $(q_c/q)^2 \bar{\beta}$ where,
\begin{equation}
    \bar{\beta} = \frac{\beta}{\beta_c} = \frac{(1 + \bar{s}\tau)^{-2} + \alpha}{1 + \alpha},
\end{equation}
and $\alpha = s_0 k_b T_{\perp}/(1+s_0)U_0$. In the limit of large damping where $v_{th}/v_b \rightarrow \infty$, we find $\tau \rightarrow 0$ and we retrieve the response,
\begin{equation}
    \bar{s} = \pm \sqrt{1/\bar{q}^2 - 1}, \ \ \tau \rightarrow 0.
\end{equation}
This result is precisely the growth increment of the Bespalov-Talanov instability except with the critical wavenumber modified by the kinetic response as in Eq. \ref{eq:critical}. To see this clearly, we define a new dimensionless longitudinal spatial frequency based not on $q$ but on $q_c$: $\hat{s} = \bar{s} \bar{q}^2$. With this new definition the growth rate can be expressed as,
\begin{equation}
    \hat{s} = \bar{q} \sqrt{1 - \bar{q}^2}.
\label{eq:besp_tal}
\end{equation}
The maximum growth-rate is then $\hat{s} = 1/2$ which occurs at $\bar{q} = 1/\sqrt{2}$. 

We interpret this limit as corresponding to a fast redistribution of the atoms in response to the dipole forces, relative to the longitudinal convection. Since we are seeking stationary solutions, this corresponds to a negligible distance in $z$ over which a given field perturbation results in an atomic response. The kinetic feedback thus presents no differently than the much faster electronic two-level nonlinear response. The latter instability was observed experimentally in back-reflection experiments by Camara and co-workers, who found a saturation threshold for self-focusing ($n_2 > 0$) and a most unstable wavenumber of $\Lambda^2 = \lambda d/(N+1/4)$ where $N$ is an integer of the same sign as the back-reflection distance $d$ \cite{Camara2015}. To compare with our result for the continuous optomechanical system we compute $\Lambda$ as,
\begin{equation}
\Lambda^2 = \frac{8\pi^2}{q_c^2} = \frac{\lambda^2}{2(n_0-1)} \frac{1+s_0}{s_0}.
\end{equation}
In contrast to the back-reflection arrangement, both the refractive index and saturation parameter select the most unstable transverse wavelength, with higher saturation and weaker refractive index pushing the instability to larger scale lengths. We can observe that the strongest opto-mechanical feedback then occurs at the scale where $V^2 \sim 1$ which is mediated by atoms for which $v \sim \tau v_{th}$.
%
%
\begin{figure}[htp]
    \centering
    \includegraphics[width=3.5in]{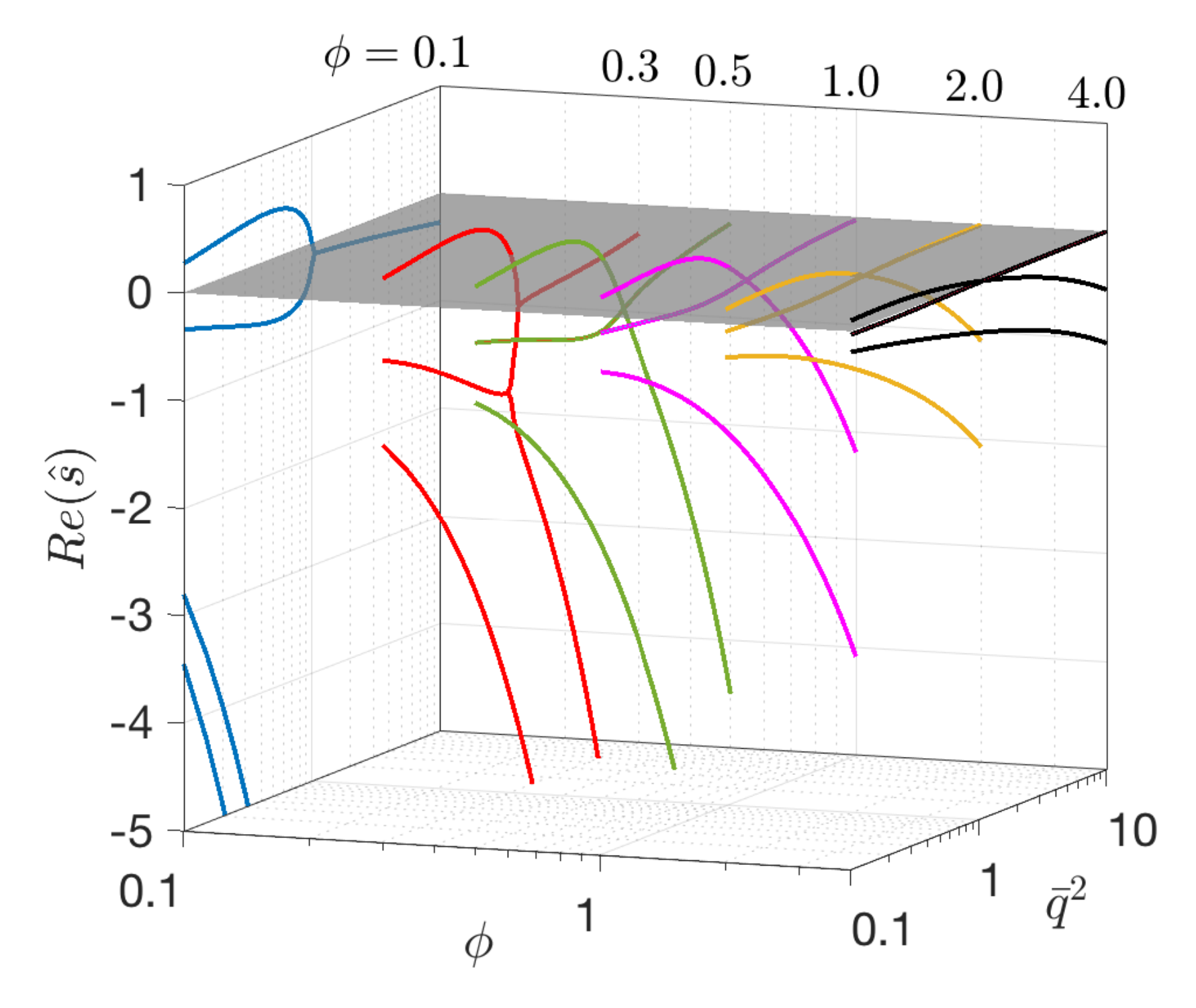}
    \caption{Real part of $\hat{s}$ corresponding to the growth rate. $Re(\hat{s}) > 0$ only for $\bar{q}>0$.}
    \label{fig:s_real}
\end{figure}

\begin{figure}[htp]
    \centering
    \includegraphics[width=3.5in]{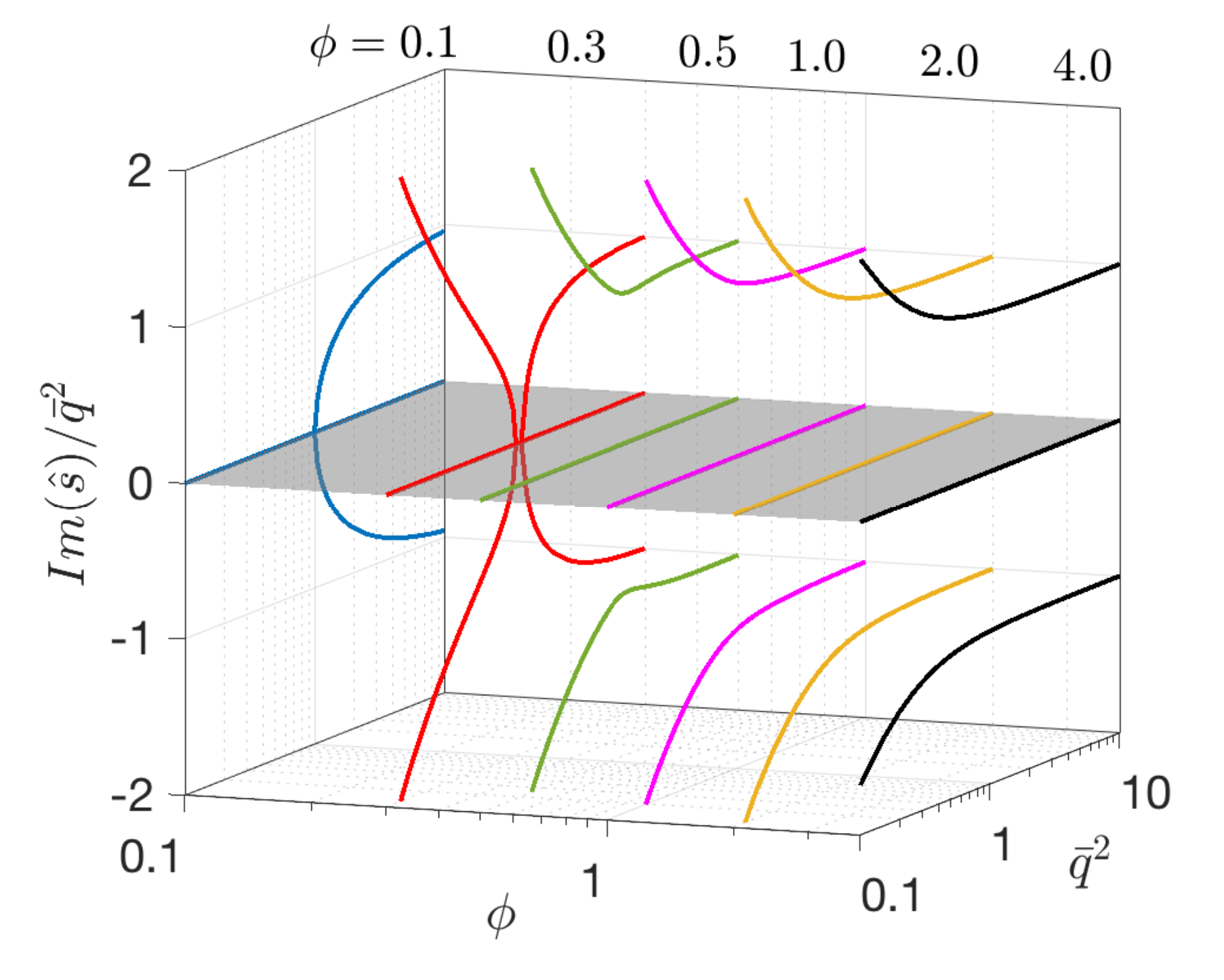}
    \caption{Imaginary part of $\hat{s}$ representing the spatial oscillation frequency. At high wavenumbers diffraction dominates and oscillation occurs at the Talbot length.}
    \label{fig:s_imag}
\end{figure}
We next examine the opposite limit, $\tau >> 1$, corresponding to a fast, cold beam $v_{th}/v_b \rightarrow 0$. In this limit we have $1 + \bar{s}\tau \rightarrow \bar{s}\tau$ from which we find $\bar{\beta} \approx \alpha / (1+\alpha)$ and $\bar{s} = \pm i \sqrt{1-\alpha/(1+\alpha)}$. For $\alpha \rightarrow 0$ we simply recover oscillation at the Talbot length. When $\alpha$ is positive (blue detuning), marginally stable oscillation occurs for any $\alpha$. However, when $\alpha$ is negative (red detuning), $|\alpha|<1$ results in oscillation while $|\alpha|>1$ can lead to strong growth or damping based on whether the saturation parameter is below or above the threshold for unconditional stabilization ($\alpha = 1$ or Eq. \ref{eq:s_thresh}).

Finally, we consider the most complex scenario where the finite redistribution time of the atoms interacts with the characteristic diffraction length. When saturation is negligible we find the dispersion in terms of the wavenumber-independent variable $\hat{s}$ and the parameter $\phi = \tau |\bar{q}|$ is given by,
\begin{equation}
    (1 + \hat{s}\phi /|\bar{q}|)^2 (\hat{s}^2 + \bar{q}^2) - 1 = 0
\label{eq:dispersion2}
\end{equation}
Eq. \ref{eq:dispersion2} is a quartic equation in $\hat{s}$ having at most four complex roots. The root locations were determined numerically as a function of $\bar{q}$ and $\phi$ and the real and imaginary parts displayed in Figs. \ref{fig:s_real} and \ref{fig:s_imag}, respectively. Notably, a single unstable root always occurs when $\bar{q}<1$ and for which $Im(\hat{s}) = 0$.
\begin{figure}[htp]
    \centering
    \includegraphics[width=3.5in]{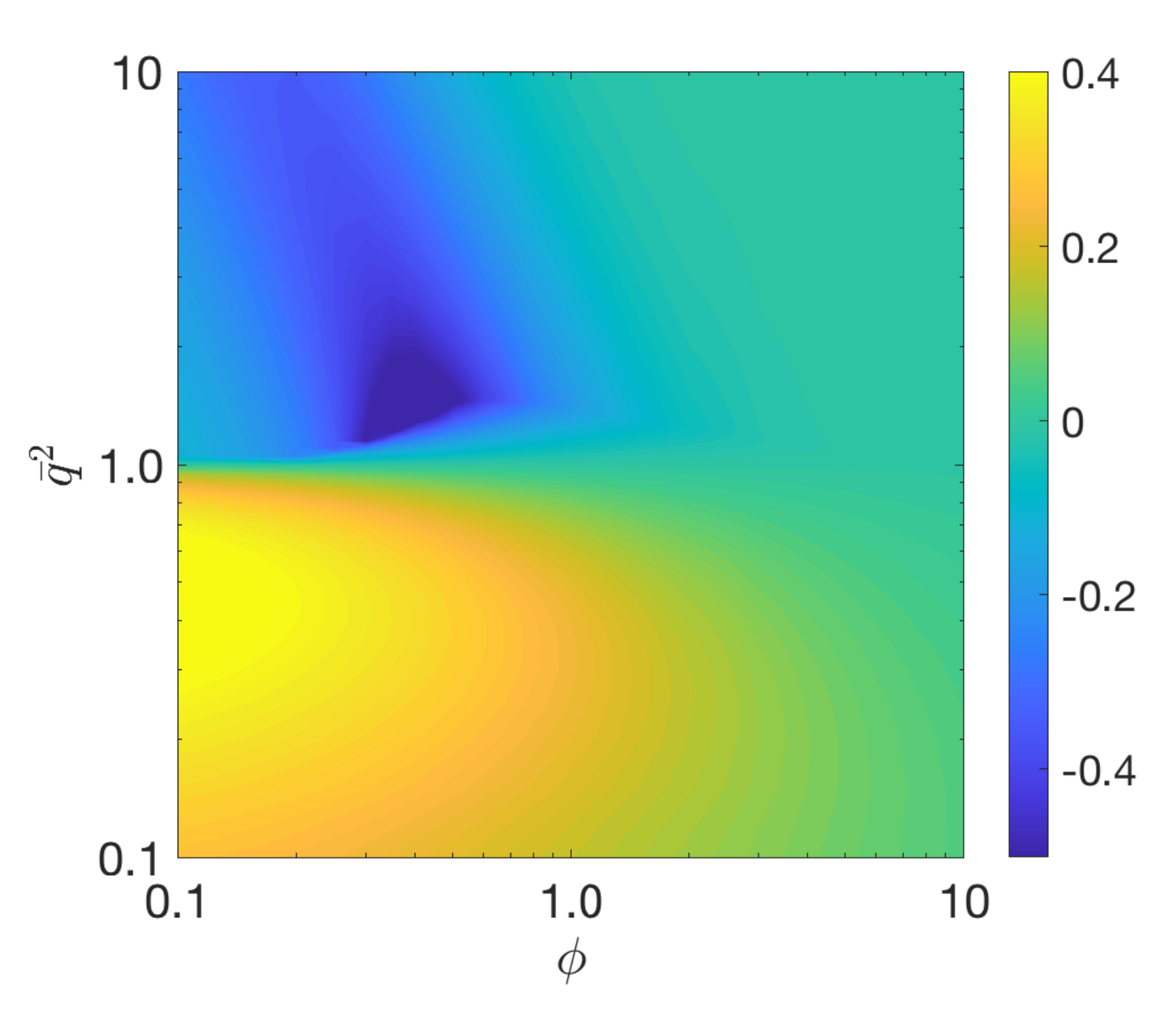}
    \caption{Maximum growth/decay rate $Max(\hat{s})$ showing strong damping for wavenumbers below $\bar{q} = 1$ for $\phi \sim 0.4$. At larger values of $\phi$ the instability growth rate is reduced due to the relatively slow transverse motion of the atoms.}
    \label{fig:growthrate}
\end{figure}

When $\bar{q}$ is large the roots one pair of roots are approximately located at $\hat{s} = \pm i\bar{q}$ while the second pair appear on the real axis in the vicinity of $\hat{s} = -|\bar{q}|/\phi$, the latter corresponding to the kinetic contribution. In this regime where $\bar{q}>>1$ the diffractive modulation again produces an oscillatory response at the Talbot length which is now weakly damped (as compared with Eq. \ref{eq:besp_tal}), due to the kinetic damping mechanism discussed in Sec. \ref{sec:damping}.   
Considering now the variation with $\phi$, we can observe that for $\phi << 1$ the roots are decoupled over a large range of spatial scales $\bar{q}$ (see Fig. \ref{fig:s_real}), since the thermal dephasing suppresses the interaction with diffraction. In the intermediate regime $\phi \sim 1$, and especially near $\phi \approx 1/3$, we observe a strong interaction between diffraction and the kinetic response. This results in the disappearance of one of the conjugate roots pairs (seen in Fig. \ref{fig:s_imag}, red curve) resulting in an overdamped within a small range of $q$ just below $q_c$. This can be seen in Fig. \ref{fig:growthrate} which shows $Max(\hat{s})$ as a function of $\bar{q}$ and $\phi$. For $\phi << 1$ the most unstable wavenumber occurs at $\bar{q}^2 = 1/2$ while shifting to lower wavenumber when $\phi$ increases as the atoms require relatively more time (also, distance) to spatially redistribute due to their lower transverse velocity.  

\vspace{-10pt}
\subsection{Landau Damping} \label{sec:damping}

The damping observed in Sec. \ref{sec:stability} is similar in nature to other forms of damping in collisionless systems first identified by Landau in electrostatic plasma waves \cite{Landau1946}. In plasmas, kinetic damping occurs due to momentum and energy exchange between the field and near-resonant particles with velocities near the phase velocity of the perturbation \cite{Gurnett2017}. In the continuous optomechanical system, the resonant particle velocity $v_p$ is related to the diffraction angle of a perturbation of wavenumber $q$ by $v_p = v_b / q L_d$. 

We examine the evolution of the velocity-space perturbation by direct integration of the Fourier transformed linearized equations which constitute a set of ordinary differential equations, one for the electric field and an infinite number in transverse velocity space. Discretizing velocity space into a finite number of bins (groups), the equations are solved by a first order implicit finite difference method centered in $z$. The density perturbation is evaluated by trapezoidal integration of the distribution function and simulations were checked for convergence and cross-verified by comparing with numerical evaluation of the inverse Laplace transform of Eq. \ref{eq:system} using the numerical method of Abate \cite{Abate2004} \footnote{Notably, the use of a finite number of velocity bins, which act as weakly coupled modes, leads to an inexact recurrence of the initial perturbation after some distance which becomes larger as the number of velocity bins is increased. Phenomena of this type were first identified by Fermi, Pasta and Ulam \cite{Fermi1955} in the context of oscillations of a finite string but in this context represent a non-physical effect of the velocity-space discretization which is to be avoided.}. 
\begin{figure}[h]
\centering
\includegraphics[width=3.25in]{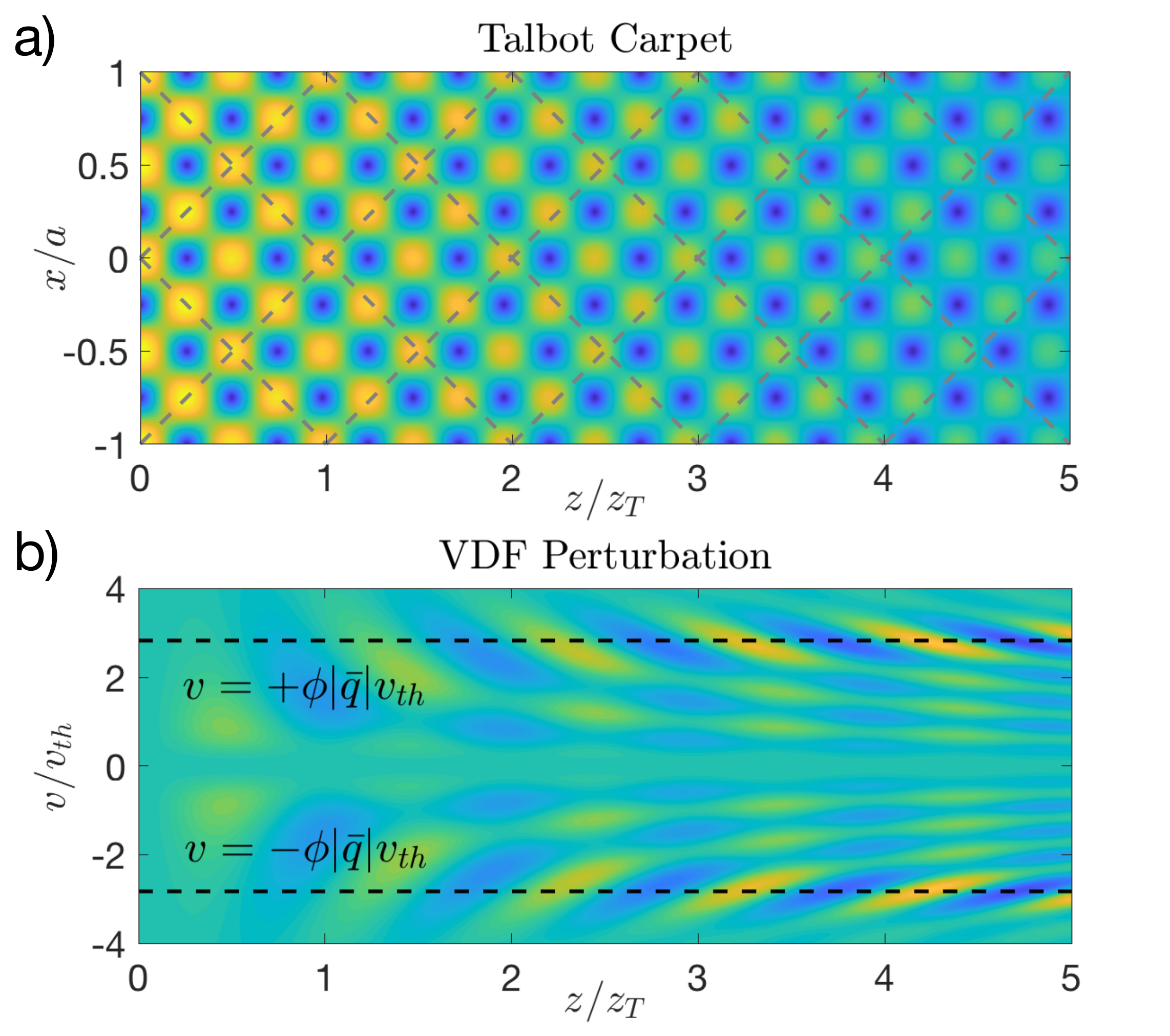}
\caption{a) Talbot carpet, and b) VDF Perturbation for $\bar{q}^2 = 2$ and $\phi = 2$. Particles with thermal velocity $\pm v/v_b$ travel along the dashed gray lines indicated in (a) with slope $\pm a/z_T$ and experience a resonant interaction. This corresponds to the condition $\tau = \bar{q}\phi = 1$.}
\label{fig:talbot_vdf}
\end{figure}

As an illustrative example of the velocity-space perturbation, results for a damped perturbation with $\bar{q}^2 = 2$ and $\phi = 2$ are shown in Fig. \ref{fig:talbot_vdf}, where an initial perturbation was launched with $\hat{E}(0) = 1$. In panel (a) we show the corresponding Talbot carpet with the transverse coordinate normalized to $a = 2\pi/q$ and the longitudinal distance normalized to the Talbot length. The dashed lines indicate the trajectory of resonant particles along preferred directions on the Talbot carpet with angles $a/z_T = \lambda/2a$. We can observe in panel (b) that the velocity-space perturbation oscillates about the velocity $v = \pm \tau v_{th}$. Notably, the shift in the effective Talbot length caused by the density perturbation appears to only temporarily trap atoms along the characteristics of Fig. \ref{fig:talbot_vdf}(a). The oscillations can be contrasted with the velocity-space perturbation induced by an imposed optical lattice, which leads to a continuous decrease in high velocity atoms \cite{Shneider2005b}. 

The damping is clearly indicated in panel (a) by the decreasing intensity of the perturbation irradiance. Figure \ref{fig:talbot_vdf} (b) shows the velocity-space perturbation which while initially zero grows within one Talbot cycle and becomes most prominent around the resonant velocity $v = \pm \phi \bar{q} v_{th}$, i.e. $\tau = 1$. Note that the commensurate density perturbation is sufficient to shift the Talbot carpet shown in Fig. \ref{fig:talbot_vdf} relative to the unperturbed case, as indicated by a careful examination of the dashed line crossings in Fig. \ref{fig:talbot_vdf} (a) relative to the intensity maxima. 

\begin{figure}[h]
\centering
\includegraphics[width=3.25in]{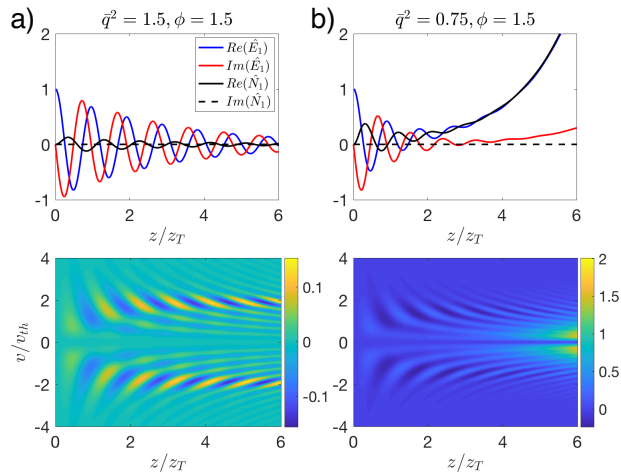}
\caption{Field (top) and kinetic (bottom) responses for selected cases exhibiting a) damping, b) growth. Both simulations are initialized with a field perturbation $\hat{E}_1(z=0) = 1$.}
\label{fig:responses}
\end{figure}

Additional calculations were performed for a range of $\bar{q}$ and $\phi$, with several representative results shown in Fig. \ref{fig:responses}. In the top panel we see the real and imaginary components of the field response again with an initial perturbation $\hat{E}(0) = 1$, where the (purely real) density perturbation is shown by a black solid line. The result for $\bar{q}^2 = 1.5$, $\phi = 1.5$ illustrates the excitation of resonant atoms near $v = \pm \tau v_{th}$ and the phase mixing that occurs as the wave is damped. 

In contrast, Fig. \ref{fig:responses} (b) shows an unstable case ($\bar{q}<1$) which illustrates the in-phase coupling of the density perturbation and real part of the field leading to runaway growth. The lower panel showing the velocity perturbation shows how the slow atoms become trapped within the Talbot carpet and accumulate as the perturbation continues to grow. 
%
%
%
\vspace{-10pt}
\subsection{Discussion} \label{sec:discussion}
%
%
%
In summary, we have derived the dispersion relation and growth rates for the modulational instability including both kinetic and electronic feedback. In particular, we show that the most unstable wavenumber occurs slightly below the critical wavenumber and is set by the combination of refractive, kinetic and electronic feedback in Eq. \ref{eq:critical}. A new dimensionless parameter $\phi$ was found which captures the delayed response of the atomic motion to the field perturbation, yielding regions of strong damping near $\phi = 1/3$. The mechanism for this collisionless damping was shown to be analogous to Landau damping and creates the strongest velocity-space perturbation near the resonant velocity $v = \pm \phi |\bar{q}| v_{th}$.

The observation of the damping and growth described in Sec.~\ref{sec:damping} should be experimentally accessible with cold atomic beams. For a precise correspondence with the above analysis, mono-energetic beams may be produced by the moving mollasses technique \cite{Berthoud1999} although the theory can be readily extended to atomic beams of finite longitudinal velocity spread. If one considers an atomic beam Rb atomic beam with $N_0 = 10^{10} $ cm$^{-3}$ and $\Gamma = 2\pi \times 6.07$ MHz detuned by $2\pi \times 100$ MHz, we obtain a refractive index of $n_0 = 1 + 5.4 \times 10^{-6}$. If the beam is cooled to a Doppler temperature of 146 $\mu$K and irradiated at $s_0 = 0.1$, the resulting critical wavenumber is $q_c = 579$ cm$^{-1}$ with diffraction length $L_d = 7.0$ mm. With a similar experiments to Bjorkholm and Pearson \cite{Bjorkholm1978,Pearson1980}, such self-structuring as seen by Labeyrie \cite{Labeyrie2014} could be observed in the continuous optomechanical interaction within a distance of several Talbot lengths, which for $q = q_c/\sqrt{2}$ could be implemented in table-top experiments.

Kinetic damping and phase mixing may also occur in contexts other than the homogeneous solution examined here, in particular optomechanical solitons. The results above show that for a sufficiently large atomic beam structures of wavenumber $\sim q_c/\sqrt{2}$ will grow in amplitude leading to filamentation. Although it is unclear whether the filaments themselves would be stable, the collisionless damping mechanism provides a potential stabilization method enabling long-range propagation through a vacuum environment. It can be expected that angular momentum will play a key role in the kinetic interaction for axisymmetric beams, particularly for the lowest order bright-bright solitons for which stationary distributions correspond to the conserved integrals of motion in the 2D central-potential. The interaction between the kinetic damping and other stabilizing processes, such as the non-local response \cite{Skupin2007} or optical angular momentum \cite{Gibson2020}, may provide further opportunities to tailor the dynamics of coupled beam propagation.       
%
%
\begin{acknowledgments}
The author wishes to acknowledge the support of the NASA Innovative Advanced Concepts Program under grant 80NSSC19K0974. The author is also grateful for discussions with K. Hara, P. Kumar and A. Castillo and comments on the manuscript in preparation by M. Shneider and H. Morgan. 
\end{acknowledgments}


\bibliography{library}

\begin{thebibliography}{40}%
\makeatletter
\providecommand \@ifxundefined [1]{%
 \@ifx{#1\undefined}
}%
\providecommand \@ifnum [1]{%
 \ifnum #1\expandafter \@firstoftwo
 \else \expandafter \@secondoftwo
 \fi
}%
\providecommand \@ifx [1]{%
 \ifx #1\expandafter \@firstoftwo
 \else \expandafter \@secondoftwo
 \fi
}%
\providecommand \natexlab [1]{#1}%
\providecommand \enquote  [1]{``#1''}%
\providecommand \bibnamefont  [1]{#1}%
\providecommand \bibfnamefont [1]{#1}%
\providecommand \citenamefont [1]{#1}%
\providecommand \href@noop [0]{\@secondoftwo}%
\providecommand \href [0]{\begingroup \@sanitize@url \@href}%
\providecommand \@href[1]{\@@startlink{#1}\@@href}%
\providecommand \@@href[1]{\endgroup#1\@@endlink}%
\providecommand \@sanitize@url [0]{\catcode `\\12\catcode `\$12\catcode
  `\&12\catcode `\#12\catcode `\^12\catcode `\_12\catcode `\%12\relax}%
\providecommand \@@startlink[1]{}%
\providecommand \@@endlink[0]{}%
\providecommand \url  [0]{\begingroup\@sanitize@url \@url }%
\providecommand \@url [1]{\endgroup\@href {#1}{\urlprefix }}%
\providecommand \urlprefix  [0]{URL }%
\providecommand \Eprint [0]{\href }%
\providecommand \doibase [0]{https://doi.org/}%
\providecommand \selectlanguage [0]{\@gobble}%
\providecommand \bibinfo  [0]{\@secondoftwo}%
\providecommand \bibfield  [0]{\@secondoftwo}%
\providecommand \translation [1]{[#1]}%
\providecommand \BibitemOpen [0]{}%
\providecommand \bibitemStop [0]{}%
\providecommand \bibitemNoStop [0]{.\EOS\space}%
\providecommand \EOS [0]{\spacefactor3000\relax}%
\providecommand \BibitemShut  [1]{\csname bibitem#1\endcsname}%
\let\auto@bib@innerbib\@empty
\bibitem [{\citenamefont {Bespalov}\ and\ \citenamefont
  {Talanov}(1966)}]{Bespalov1966}%
  \BibitemOpen
  \bibfield  {author} {\bibinfo {author} {\bibfnamefont {V.~I.}\ \bibnamefont
  {Bespalov}}\ and\ \bibinfo {author} {\bibfnamefont {V.~I.}\ \bibnamefont
  {Talanov}},\ }\bibfield  {title} {\bibinfo {title} {{Filamentary Structure of
  Light Beams in Nonlinear Media}},\ }\href@noop {} {\bibfield  {journal}
  {\bibinfo  {journal} {Pis'ma Zh. Eksp. Teor. Fiz. [JETP Lett.]}\ }\textbf
  {\bibinfo {volume} {3}} (\bibinfo {year} {1966})}\BibitemShut {NoStop}%
\bibitem [{\citenamefont {Tesio}\ \emph {et~al.}(2013)\citenamefont {Tesio},
  \citenamefont {Robb}, \citenamefont {Ackemann}, \citenamefont {Firth},\ and\
  \citenamefont {Oppo}}]{Tesio2013}%
  \BibitemOpen
  \bibfield  {author} {\bibinfo {author} {\bibfnamefont {E.}~\bibnamefont
  {Tesio}}, \bibinfo {author} {\bibfnamefont {G.~R.~M.}\ \bibnamefont {Robb}},
  \bibinfo {author} {\bibfnamefont {T.}~\bibnamefont {Ackemann}}, \bibinfo
  {author} {\bibfnamefont {W.~J.}\ \bibnamefont {Firth}},\ and\ \bibinfo
  {author} {\bibfnamefont {G.}~\bibnamefont {Oppo}},\ }\bibfield  {title}
  {\bibinfo {title} {{Dissipative solitons in the coupled dynamics of light and
  cold atoms}},\ }\href {https://doi.org/10.1364/OE.21.026144} {\bibfield
  {journal} {\bibinfo  {journal} {Optics Express}\ }\textbf {\bibinfo {volume}
  {21}},\ \bibinfo {pages} {26144} (\bibinfo {year} {2013})},\ \Eprint
  {https://arxiv.org/abs/arXiv:1308.1226} {arXiv:arXiv:1308.1226} \BibitemShut
  {NoStop}%
\bibitem [{\citenamefont {Tesio}\ \emph {et~al.}(2014)\citenamefont {Tesio},
  \citenamefont {Robb}, \citenamefont {Ackemann}, \citenamefont {Firth},\ and\
  \citenamefont {Oppo}}]{Tesio2014}%
  \BibitemOpen
  \bibfield  {author} {\bibinfo {author} {\bibfnamefont {E.}~\bibnamefont
  {Tesio}}, \bibinfo {author} {\bibfnamefont {G.~R.~M.}\ \bibnamefont {Robb}},
  \bibinfo {author} {\bibfnamefont {T.}~\bibnamefont {Ackemann}}, \bibinfo
  {author} {\bibfnamefont {W.~J.}\ \bibnamefont {Firth}},\ and\ \bibinfo
  {author} {\bibfnamefont {G.}~\bibnamefont {Oppo}},\ }\bibfield  {title}
  {\bibinfo {title} {{Kinetic Theory for Transverse Optomechanical
  Instabilities}},\ }\href {https://doi.org/10.1103/PhysRevLett.112.043901}
  {\bibfield  {journal} {\bibinfo  {journal} {Physical Review Letters}\
  }\textbf {\bibinfo {volume} {112}},\ \bibinfo {pages} {1} (\bibinfo {year}
  {2014})}\BibitemShut {NoStop}%
\bibitem [{\citenamefont {Labeyrie}\ \emph {et~al.}(2014)\citenamefont
  {Labeyrie}, \citenamefont {Tesio}, \citenamefont {Gomes}, \citenamefont
  {Oppo}, \citenamefont {Firth}, \citenamefont {Robb}, \citenamefont {Arnold},
  \citenamefont {Kaiser},\ and\ \citenamefont {Ackemann}}]{Labeyrie2014}%
  \BibitemOpen
  \bibfield  {author} {\bibinfo {author} {\bibfnamefont {G.}~\bibnamefont
  {Labeyrie}}, \bibinfo {author} {\bibfnamefont {E.}~\bibnamefont {Tesio}},
  \bibinfo {author} {\bibfnamefont {P.~M.}\ \bibnamefont {Gomes}}, \bibinfo
  {author} {\bibfnamefont {G.}~\bibnamefont {Oppo}}, \bibinfo {author}
  {\bibfnamefont {W.~J.}\ \bibnamefont {Firth}}, \bibinfo {author}
  {\bibfnamefont {G.~R.~M.}\ \bibnamefont {Robb}}, \bibinfo {author}
  {\bibfnamefont {A.~S.}\ \bibnamefont {Arnold}}, \bibinfo {author}
  {\bibfnamefont {R.}~\bibnamefont {Kaiser}},\ and\ \bibinfo {author}
  {\bibfnamefont {T.}~\bibnamefont {Ackemann}},\ }\bibfield  {title} {\bibinfo
  {title} {{Optomechanical self-structuring in a cold atomic gas}},\ }\href
  {https://doi.org/10.1038/nphoton.2014.52} {\bibfield  {journal} {\bibinfo
  {journal} {Nature Photonics}\ }\textbf {\bibinfo {volume} {8}},\ \bibinfo
  {pages} {321} (\bibinfo {year} {2014})}\BibitemShut {NoStop}%
\bibitem [{\citenamefont {Baio}\ \emph {et~al.}(2020)\citenamefont {Baio},
  \citenamefont {Robb}, \citenamefont {Yao},\ and\ \citenamefont
  {Oppo}}]{Baio2020}%
  \BibitemOpen
  \bibfield  {author} {\bibinfo {author} {\bibfnamefont {G.}~\bibnamefont
  {Baio}}, \bibinfo {author} {\bibfnamefont {G.~R.}\ \bibnamefont {Robb}},
  \bibinfo {author} {\bibfnamefont {A.~M.}\ \bibnamefont {Yao}},\ and\ \bibinfo
  {author} {\bibfnamefont {G.~L.}\ \bibnamefont {Oppo}},\ }\bibfield  {title}
  {\bibinfo {title} {{Optomechanical transport of cold atoms induced by
  structured light}},\ }\bibfield  {journal} {\bibinfo  {journal} {arXiv}\
  }\textbf {\bibinfo {volume} {023126}},\ \href
  {https://doi.org/10.1103/physrevresearch.2.023126}
  {10.1103/physrevresearch.2.023126} (\bibinfo {year} {2020}),\ \Eprint
  {https://arxiv.org/abs/2001.04892} {arXiv:2001.04892} \BibitemShut {NoStop}%
\bibitem [{\citenamefont {Schmittberger}\ and\ \citenamefont
  {Gauthier}(2016)}]{Schmittberger2016a}%
  \BibitemOpen
  \bibfield  {author} {\bibinfo {author} {\bibfnamefont {B.~L.}\ \bibnamefont
  {Schmittberger}}\ and\ \bibinfo {author} {\bibfnamefont {D.~J.}\ \bibnamefont
  {Gauthier}},\ }\bibfield  {title} {\bibinfo {title} {{Spontaneous emergence
  of free-space optical and atomic patterns}},\ }\bibfield  {journal} {\bibinfo
   {journal} {New Journal of Physics}\ }\textbf {\bibinfo {volume} {18}},\
  \href {https://doi.org/10.1088/1367-2630/18/10/103021}
  {10.1088/1367-2630/18/10/103021} (\bibinfo {year} {2016})\BibitemShut
  {NoStop}%
\bibitem [{\citenamefont {Camara}\ \emph {et~al.}(2015)\citenamefont {Camara},
  \citenamefont {Kaiser}, \citenamefont {Labeyrie}, \citenamefont {Firth},
  \citenamefont {Oppo}, \citenamefont {Robb}, \citenamefont {Arnold},\ and\
  \citenamefont {Ackemann}}]{Camara2015}%
  \BibitemOpen
  \bibfield  {author} {\bibinfo {author} {\bibfnamefont {A.}~\bibnamefont
  {Camara}}, \bibinfo {author} {\bibfnamefont {R.}~\bibnamefont {Kaiser}},
  \bibinfo {author} {\bibfnamefont {G.}~\bibnamefont {Labeyrie}}, \bibinfo
  {author} {\bibfnamefont {W.~J.}\ \bibnamefont {Firth}}, \bibinfo {author}
  {\bibfnamefont {G.~L.}\ \bibnamefont {Oppo}}, \bibinfo {author}
  {\bibfnamefont {G.~R.}\ \bibnamefont {Robb}}, \bibinfo {author}
  {\bibfnamefont {A.~S.}\ \bibnamefont {Arnold}},\ and\ \bibinfo {author}
  {\bibfnamefont {T.}~\bibnamefont {Ackemann}},\ }\bibfield  {title} {\bibinfo
  {title} {{Optical pattern formation with a two-level nonlinearity}},\ }\href
  {https://doi.org/10.1103/PhysRevA.92.013820} {\bibfield  {journal} {\bibinfo
  {journal} {Physical Review A - Atomic, Molecular, and Optical Physics}\
  }\textbf {\bibinfo {volume} {92}},\ \bibinfo {pages} {1} (\bibinfo {year}
  {2015})},\ \Eprint {https://arxiv.org/abs/1506.06638} {arXiv:1506.06638}
  \BibitemShut {NoStop}%
\bibitem [{\citenamefont {Khaykovich}\ \emph {et~al.}(2002)\citenamefont
  {Khaykovich}, \citenamefont {Schreck}, \citenamefont {Ferrari}, \citenamefont
  {Bourdel}, \citenamefont {Cubizolles}, \citenamefont {Carr}, \citenamefont
  {Castin},\ and\ \citenamefont {Salomon}}]{Khaykovich2002}%
  \BibitemOpen
  \bibfield  {author} {\bibinfo {author} {\bibfnamefont {L.}~\bibnamefont
  {Khaykovich}}, \bibinfo {author} {\bibfnamefont {F.}~\bibnamefont {Schreck}},
  \bibinfo {author} {\bibfnamefont {G.}~\bibnamefont {Ferrari}}, \bibinfo
  {author} {\bibfnamefont {T.}~\bibnamefont {Bourdel}}, \bibinfo {author}
  {\bibfnamefont {J.}~\bibnamefont {Cubizolles}}, \bibinfo {author}
  {\bibfnamefont {L.~D.}\ \bibnamefont {Carr}}, \bibinfo {author}
  {\bibfnamefont {Y.}~\bibnamefont {Castin}},\ and\ \bibinfo {author}
  {\bibfnamefont {C.}~\bibnamefont {Salomon}},\ }\bibfield  {title} {\bibinfo
  {title} {{Formation of a matter-wave bright soliton}},\ }\href
  {https://doi.org/10.1126/science.1071021} {\bibfield  {journal} {\bibinfo
  {journal} {Science}\ }\textbf {\bibinfo {volume} {296}},\ \bibinfo {pages}
  {1290} (\bibinfo {year} {2002})},\ \Eprint {https://arxiv.org/abs/0205378}
  {arXiv:0205378 [cond-mat]} \BibitemShut {NoStop}%
\bibitem [{\citenamefont {Saffman}(1998)}]{Saffman1998}%
  \BibitemOpen
  \bibfield  {author} {\bibinfo {author} {\bibfnamefont {M.}~\bibnamefont
  {Saffman}},\ }\bibfield  {title} {\bibinfo {title} {{Self-Induced Dipole
  Force and Filamentation Instability of a Matter Wave}},\ }\href@noop {}
  {\bibfield  {journal} {\bibinfo  {journal} {Physical Review Letters}\
  }\textbf {\bibinfo {volume} {81}},\ \bibinfo {pages} {65} (\bibinfo {year}
  {1998})}\BibitemShut {NoStop}%
\bibitem [{\citenamefont {Saffman}\ and\ \citenamefont
  {Skryabin}(2001)}]{Saffman2001}%
  \BibitemOpen
  \bibfield  {author} {\bibinfo {author} {\bibfnamefont {M.}~\bibnamefont
  {Saffman}}\ and\ \bibinfo {author} {\bibfnamefont {D.~V.}\ \bibnamefont
  {Skryabin}},\ }\bibfield  {title} {\bibinfo {title} {{Spatial Solitons}},\
  }in\ \href@noop {} {\emph {\bibinfo {booktitle} {Spatial Solitons}}},\
  \bibinfo {editor} {edited by\ \bibinfo {editor} {\bibfnamefont {W.~T.}\
  \bibnamefont {Rhodes}}}\ (\bibinfo  {publisher} {Springer-Verlag},\ \bibinfo
  {address} {Berlin Heidelberg},\ \bibinfo {year} {2001})\ Chap.\ \bibinfo
  {chapter} {Coupled Pr}\BibitemShut {NoStop}%
\bibitem [{\citenamefont {Askar'yan}(1962)}]{Askaryan1962}%
  \BibitemOpen
  \bibfield  {author} {\bibinfo {author} {\bibfnamefont {G.~A.}\ \bibnamefont
  {Askar'yan}},\ }\bibfield  {title} {\bibinfo {title} {{Effects of the
  Gradient of a Strong Electromagnetic Beam on Electrons and Atoms}},\
  }\href@noop {} {\bibfield  {journal} {\bibinfo  {journal} {Soviet Physics
  JETP}\ ,\ \bibinfo {pages} {1088}} (\bibinfo {year} {1962})}\BibitemShut
  {NoStop}%
\bibitem [{\citenamefont {Bjorkholm}\ \emph {et~al.}(1978)\citenamefont
  {Bjorkholm}, \citenamefont {Freeman}, \citenamefont {Ashkin},\ and\
  \citenamefont {Pearson}}]{Bjorkholm1978}%
  \BibitemOpen
  \bibfield  {author} {\bibinfo {author} {\bibfnamefont {J.~E.}\ \bibnamefont
  {Bjorkholm}}, \bibinfo {author} {\bibfnamefont {R.~R.}\ \bibnamefont
  {Freeman}}, \bibinfo {author} {\bibfnamefont {A.}~\bibnamefont {Ashkin}},\
  and\ \bibinfo {author} {\bibfnamefont {D.~B.}\ \bibnamefont {Pearson}},\
  }\bibfield  {title} {\bibinfo {title} {{Observation of Focusing of Neutral
  Atoms by the Dipole Forces of Resonance-Radiation Pressure}},\ }\href@noop {}
  {\bibfield  {journal} {\bibinfo  {journal} {Physical Review Letters}\
  }\textbf {\bibinfo {volume} {41}},\ \bibinfo {pages} {1361} (\bibinfo {year}
  {1978})}\BibitemShut {NoStop}%
\bibitem [{\citenamefont {Bjorkholm}\ \emph {et~al.}(1980)\citenamefont
  {Bjorkholm}, \citenamefont {Freeman}, \citenamefont {Ashkin},\ and\
  \citenamefont {Pearson}}]{Bjorkholm1980}%
  \BibitemOpen
  \bibfield  {author} {\bibinfo {author} {\bibfnamefont {E.}~\bibnamefont
  {Bjorkholm}}, \bibinfo {author} {\bibfnamefont {R.~R.}\ \bibnamefont
  {Freeman}}, \bibinfo {author} {\bibfnamefont {A.}~\bibnamefont {Ashkin}},\
  and\ \bibinfo {author} {\bibfnamefont {D.~B.}\ \bibnamefont {Pearson}},\
  }\bibfield  {title} {\bibinfo {title} {{Experimental observation of the
  influence of the quantum fluctuations of resonance-radiation pressure}},\
  }\href@noop {} {\bibfield  {journal} {\bibinfo  {journal} {Optics Letters}\
  }\textbf {\bibinfo {volume} {5}},\ \bibinfo {pages} {111} (\bibinfo {year}
  {1980})}\BibitemShut {NoStop}%
\bibitem [{\citenamefont {Pearson}\ \emph {et~al.}(1980)\citenamefont
  {Pearson}, \citenamefont {Freeman}, \citenamefont {Bjorkholm},\ and\
  \citenamefont {Ashkin}}]{Pearson1980}%
  \BibitemOpen
  \bibfield  {author} {\bibinfo {author} {\bibfnamefont {D.~B.}\ \bibnamefont
  {Pearson}}, \bibinfo {author} {\bibfnamefont {R.~R.}\ \bibnamefont
  {Freeman}}, \bibinfo {author} {\bibfnamefont {J.~E.}\ \bibnamefont
  {Bjorkholm}},\ and\ \bibinfo {author} {\bibfnamefont {A.}~\bibnamefont
  {Ashkin}},\ }\bibfield  {title} {\bibinfo {title} {{Focusing and defocusing
  of neutral atomic beams using resonance-radiation pressure}},\ }\href
  {https://doi.org/10.1063/1.91289} {\bibfield  {journal} {\bibinfo  {journal}
  {Applied Physics Letters}\ }\textbf {\bibinfo {volume} {36}},\ \bibinfo
  {pages} {99} (\bibinfo {year} {1980})}\BibitemShut {NoStop}%
\bibitem [{\citenamefont {Helseth}(2002)}]{Helseth2002}%
  \BibitemOpen
  \bibfield  {author} {\bibinfo {author} {\bibfnamefont {L.~E.}\ \bibnamefont
  {Helseth}},\ }\bibfield  {title} {\bibinfo {title} {{Focusing of atoms with
  spatially localized light pulses}},\ }\bibfield  {journal} {\bibinfo
  {journal} {Physical Review A - Atomic, Molecular, and Optical Physics}\
  }\textbf {\bibinfo {volume} {66}},\ \href
  {https://doi.org/10.1103/PhysRevA.66.053609} {10.1103/PhysRevA.66.053609}
  (\bibinfo {year} {2002})\BibitemShut {NoStop}%
\bibitem [{\citenamefont {Riis}\ \emph {et~al.}(1990)\citenamefont {Riis},
  \citenamefont {Weiss}, \citenamefont {Moler},\ and\ \citenamefont
  {Chu.}}]{Riis1990}%
  \BibitemOpen
  \bibfield  {author} {\bibinfo {author} {\bibfnamefont {E.}~\bibnamefont
  {Riis}}, \bibinfo {author} {\bibfnamefont {D.}~\bibnamefont {Weiss}},
  \bibinfo {author} {\bibfnamefont {K.}~\bibnamefont {Moler}},\ and\ \bibinfo
  {author} {\bibfnamefont {S.}~\bibnamefont {Chu.}},\ }\bibfield  {title}
  {\bibinfo {title} {{Atom Fun- nel for the Production of a Slow, High-Density
  Atomic- Beam.}},\ }\href@noop {} {\bibfield  {journal} {\bibinfo  {journal}
  {Physical Review Letters}\ }\textbf {\bibinfo {volume} {64}},\ \bibinfo
  {pages} {1658} (\bibinfo {year} {1990})}\BibitemShut {NoStop}%
\bibitem [{\citenamefont {Arlt}\ \emph {et~al.}(2000)\citenamefont {Arlt},
  \citenamefont {Hitomi},\ and\ \citenamefont {Dholakia}}]{Arlt2000}%
  \BibitemOpen
  \bibfield  {author} {\bibinfo {author} {\bibfnamefont {J.}~\bibnamefont
  {Arlt}}, \bibinfo {author} {\bibfnamefont {T.}~\bibnamefont {Hitomi}},\ and\
  \bibinfo {author} {\bibfnamefont {K.}~\bibnamefont {Dholakia}},\ }\bibfield
  {title} {\bibinfo {title} {{Atom guiding along Laguerre-Gaussian and Bessel
  light beams}},\ }\href@noop {} {\bibfield  {journal} {\bibinfo  {journal}
  {Applied Physics B}\ }\textbf {\bibinfo {volume} {554}},\ \bibinfo {pages}
  {549} (\bibinfo {year} {2000})}\BibitemShut {NoStop}%
\bibitem [{\citenamefont {Song}\ \emph {et~al.}(1999)\citenamefont {Song},
  \citenamefont {Milam},\ and\ \citenamefont {Iii}}]{Song1999}%
  \BibitemOpen
  \bibfield  {author} {\bibinfo {author} {\bibfnamefont {Y.}~\bibnamefont
  {Song}}, \bibinfo {author} {\bibfnamefont {D.}~\bibnamefont {Milam}},\ and\
  \bibinfo {author} {\bibfnamefont {W.~T.~H.}\ \bibnamefont {Iii}},\ }\bibfield
   {title} {\bibinfo {title} {{Long , narrow all-light atom guide}},\
  }\href@noop {} {\bibfield  {journal} {\bibinfo  {journal} {Optics Letters}\
  }\textbf {\bibinfo {volume} {24}},\ \bibinfo {pages} {1805} (\bibinfo {year}
  {1999})}\BibitemShut {NoStop}%
\bibitem [{\citenamefont {Kumar}\ \emph {et~al.}(2021)\citenamefont {Kumar},
  \citenamefont {Kuldinow}, \citenamefont {Castillo}, \citenamefont {Gerakis},\
  and\ \citenamefont {Hara}}]{Kumar2021a}%
  \BibitemOpen
  \bibfield  {author} {\bibinfo {author} {\bibfnamefont {P.}~\bibnamefont
  {Kumar}}, \bibinfo {author} {\bibfnamefont {D.}~\bibnamefont {Kuldinow}},
  \bibinfo {author} {\bibfnamefont {A.}~\bibnamefont {Castillo}}, \bibinfo
  {author} {\bibfnamefont {A.}~\bibnamefont {Gerakis}},\ and\ \bibinfo {author}
  {\bibfnamefont {K.}~\bibnamefont {Hara}},\ }\bibfield  {title} {\bibinfo
  {title} {{Nonlinear dynamics of coupled light and particle beam
  propagation}},\ }\href {https://doi.org/10.1103/PhysRevA.103.043502}
  {\bibfield  {journal} {\bibinfo  {journal} {Physical Review A}\ }\textbf
  {\bibinfo {volume} {103}},\ \bibinfo {pages} {43502} (\bibinfo {year}
  {2021})}\BibitemShut {NoStop}%
\bibitem [{\citenamefont {{Yu.L. Klimontovich and S.N.
  Luzgin}}(1979)}]{Klimontovich1979}%
  \BibitemOpen
  \bibfield  {author} {\bibinfo {author} {\bibnamefont {{Yu.L. Klimontovich and
  S.N. Luzgin}}},\ }\bibfield  {title} {\bibinfo {title} {{The possibility of
  combined self-focusing of atomic and light beams}},\ }\href@noop {}
  {\bibfield  {journal} {\bibinfo  {journal} {Pis'ma Zh. Eksp. Teor. Fiz.}\
  }\textbf {\bibinfo {volume} {30}} (\bibinfo {year} {1979})}\BibitemShut
  {NoStop}%
\bibitem [{\citenamefont {Foot}(2005)}]{Foot2005}%
  \BibitemOpen
  \bibfield  {author} {\bibinfo {author} {\bibfnamefont {C.}~\bibnamefont
  {Foot}},\ }\href@noop {} {\emph {\bibinfo {title} {{Atomic Physics}}}}\
  (\bibinfo  {publisher} {Oxford University Press},\ \bibinfo {address}
  {Oxford},\ \bibinfo {year} {2005})\BibitemShut {NoStop}%
\bibitem [{\citenamefont {Dalibard}\ and\ \citenamefont
  {Cohen-Tannoudji}(1985)}]{Dalibard1985a}%
  \BibitemOpen
  \bibfield  {author} {\bibinfo {author} {\bibfnamefont {J.}~\bibnamefont
  {Dalibard}}\ and\ \bibinfo {author} {\bibfnamefont {C.}~\bibnamefont
  {Cohen-Tannoudji}},\ }\bibfield  {title} {\bibinfo {title} {{Atomic motion in
  laser light: Connection between semiclassical and quantum descriptions}},\
  }\href {https://doi.org/10.1088/0022-3700/18/8/019} {\bibfield  {journal}
  {\bibinfo  {journal} {Journal of Physics B: Atomic and Molecular Physics}\
  }\textbf {\bibinfo {volume} {18}},\ \bibinfo {pages} {1661} (\bibinfo {year}
  {1985})}\BibitemShut {NoStop}%
\bibitem [{\citenamefont {Kong}(1986)}]{Kong1986}%
  \BibitemOpen
  \bibfield  {author} {\bibinfo {author} {\bibfnamefont {J.~A.}\ \bibnamefont
  {Kong}},\ }\href@noop {} {\emph {\bibinfo {title} {{Electromagnetic Wave
  Theory}}}},\ \bibinfo {edition} {2nd}\ ed.\ (\bibinfo  {publisher} {Wiley},\
  \bibinfo {year} {1986})\BibitemShut {NoStop}%
\bibitem [{\citenamefont {Born}\ and\ \citenamefont {Wolf}(2000)}]{Born2000}%
  \BibitemOpen
  \bibfield  {author} {\bibinfo {author} {\bibfnamefont {M.}~\bibnamefont
  {Born}}\ and\ \bibinfo {author} {\bibfnamefont {E.}~\bibnamefont {Wolf}},\
  }\href {https://books.google.com/books?id=oV80AAAAIAAJ{\&}pgis=1} {\emph
  {\bibinfo {title} {{Principles of Optics: Electromagnetic Theory of
  Propagation, Interference and Diffraction of Light}}}}\ (\bibinfo
  {publisher} {CUP Archive},\ \bibinfo {year} {2000})\ p.\ \bibinfo {pages}
  {986}\BibitemShut {NoStop}%
\bibitem [{\citenamefont {Boyd}\ \emph {et~al.}(2009)\citenamefont {Boyd},
  \citenamefont {Svetlana},\ and\ \citenamefont {Shen}}]{Boyd2009}%
  \BibitemOpen
  \bibfield  {author} {\bibinfo {author} {\bibfnamefont {R.~W.}\ \bibnamefont
  {Boyd}}, \bibinfo {author} {\bibfnamefont {G.~L.}\ \bibnamefont {Svetlana}},\
  and\ \bibinfo {author} {\bibfnamefont {Y.~R.}\ \bibnamefont {Shen}},\ }\href
  {https://doi.org/10.1007/978-0-387-34727-1} {\emph {\bibinfo {title}
  {{Self-focusing: Past and Present}}}},\ Vol.\ \bibinfo {volume} {114}\
  (\bibinfo  {publisher} {Springer},\ \bibinfo {address} {New York, NY},\
  \bibinfo {year} {2009})\BibitemShut {NoStop}%
\bibitem [{Note1()}]{Note1}%
  \BibitemOpen
  \bibinfo {note} {In contrast to Eq. \ref {eq:paraxial}, the density
  perturbation does not couple to the gas-kinetics because the macroscopic and
  microscopic fields are approximately equal in a dilute medium \cite
  {Milonni2010}, that is $s \propto I \propto n |E|^2 \approx n_0
  |E|^2$.}\BibitemShut {Stop}%
\bibitem [{\citenamefont {Cowley}(1995)}]{Cowley1995}%
  \BibitemOpen
  \bibfield  {author} {\bibinfo {author} {\bibfnamefont {J.}~\bibnamefont
  {Cowley}},\ }\href@noop {} {\emph {\bibinfo {title} {{Diffraction
  Physics}}}}\ (\bibinfo  {publisher} {North-Holland},\ \bibinfo {year}
  {1995})\BibitemShut {NoStop}%
\bibitem [{\citenamefont {Tyson}(2000)}]{Tyson2000}%
  \BibitemOpen
  \bibfield  {author} {\bibinfo {author} {\bibfnamefont {R.~K.}\ \bibnamefont
  {Tyson}},\ }\href {https://spie.org/Publications/Book/358220?SSO=1} {\emph
  {\bibinfo {title} {{Introduction to Adaptive Optics}}}}\ (\bibinfo
  {publisher} {SPIE Press},\ \bibinfo {year} {2000})\ p.\ \bibinfo {pages}
  {130}\BibitemShut {NoStop}%
\bibitem [{\citenamefont {Saleh}\ and\ \citenamefont
  {Teich}(1991)}]{Saleh1991}%
  \BibitemOpen
  \bibfield  {author} {\bibinfo {author} {\bibfnamefont {B.~E.~A.}\
  \bibnamefont {Saleh}}\ and\ \bibinfo {author} {\bibfnamefont {M.~C.}\
  \bibnamefont {Teich}},\ }\href@noop {} {\emph {\bibinfo {title}
  {{Introduction to Photonics}}}}\ (\bibinfo  {publisher} {John Wiley {\&}
  Sons, Inc.},\ \bibinfo {year} {1991})\BibitemShut {NoStop}%
\bibitem [{\citenamefont {Boyd}(2013)}]{Boyd2013}%
  \BibitemOpen
  \bibfield  {author} {\bibinfo {author} {\bibfnamefont {R.~W.}\ \bibnamefont
  {Boyd}},\ }\href
  {https://books.google.com/books?id={\_}YpGBQAAQBAJ{\&}pgis=1} {\emph
  {\bibinfo {title} {{Nonlinear Optics}}}}\ (\bibinfo  {publisher} {Elsevier
  Science},\ \bibinfo {year} {2013})\ p.\ \bibinfo {pages} {456}\BibitemShut
  {NoStop}%
\bibitem [{\citenamefont {Gurnett}\ and\ \citenamefont
  {Bhattacharjee}(2017)}]{Gurnett2017}%
  \BibitemOpen
  \bibfield  {author} {\bibinfo {author} {\bibfnamefont {D.~A.}\ \bibnamefont
  {Gurnett}}\ and\ \bibinfo {author} {\bibfnamefont {A.}~\bibnamefont
  {Bhattacharjee}},\ }\href {https://doi.org/10.1017/9781139226059} {\emph
  {\bibinfo {title} {{Introduction to Plasma Physics}}}}\ (\bibinfo
  {publisher} {Cambridge University Press},\ \bibinfo {address} {Cambridge},\
  \bibinfo {year} {2017})\BibitemShut {NoStop}%
\bibitem [{\citenamefont {Landau}(1946)}]{Landau1946}%
  \BibitemOpen
  \bibfield  {author} {\bibinfo {author} {\bibfnamefont {L.}~\bibnamefont
  {Landau}},\ }\bibfield  {title} {\bibinfo {title} {, 10 (1946) 26.},\
  }\href@noop {} {\bibfield  {journal} {\bibinfo  {journal} {J. Phys. USSR}\
  }\textbf {\bibinfo {volume} {10}},\ \bibinfo {pages} {26} (\bibinfo {year}
  {1946})}\BibitemShut {NoStop}%
\bibitem [{\citenamefont {{Abate, J. and Valk{\'{o}}}}(2004)}]{Abate2004}%
  \BibitemOpen
  \bibfield  {author} {\bibinfo {author} {\bibfnamefont {P.}~\bibnamefont
  {{Abate, J. and Valk{\'{o}}}}},\ }\bibfield  {title} {\bibinfo {title}
  {{Multi-precision Laplace transform inversion}},\ }\href@noop {} {\bibfield
  {journal} {\bibinfo  {journal} {International Journal for Numerical Methods
  in Engineering}\ }\textbf {\bibinfo {volume} {60}},\ \bibinfo {pages} {979}
  (\bibinfo {year} {2004})}\BibitemShut {NoStop}%
\bibitem [{Note2()}]{Note2}%
  \BibitemOpen
  \bibinfo {note} {Notably, the use of a finite number of velocity bins, which
  act as weakly coupled modes, leads to an inexact recurrence of the initial
  perturbation after some distance which becomes larger as the number of
  velocity bins is increased. Phenomena of this type were first identified by
  Fermi, Pasta and Ulam \cite {Fermi1955} in the context of oscillations of a
  finite string but in this context represent a non-physical effect of the
  velocity-space discretization which is to be avoided.}\BibitemShut {Stop}%
\bibitem [{\citenamefont {Shneider}\ and\ \citenamefont
  {Barker}(2005)}]{Shneider2005b}%
  \BibitemOpen
  \bibfield  {author} {\bibinfo {author} {\bibfnamefont {M.~N.}\ \bibnamefont
  {Shneider}}\ and\ \bibinfo {author} {\bibfnamefont {P.~F.}\ \bibnamefont
  {Barker}},\ }\bibfield  {title} {\bibinfo {title} {{Optical Landau
  Damping}},\ }\href@noop {} {\bibfield  {journal} {\bibinfo  {journal}
  {Physical Review A - Atomic, Molecular, and Optical Physics}\ }\textbf
  {\bibinfo {volume} {71}},\ \bibinfo {pages} {053403} (\bibinfo {year}
  {2005})}\BibitemShut {NoStop}%
\bibitem [{\citenamefont {Berthoud}\ \emph {et~al.}(2000)\citenamefont
  {Berthoud}, \citenamefont {Fretel},\ and\ \citenamefont
  {Thomann}}]{Berthoud1999}%
  \BibitemOpen
  \bibfield  {author} {\bibinfo {author} {\bibfnamefont {P.}~\bibnamefont
  {Berthoud}}, \bibinfo {author} {\bibfnamefont {E.}~\bibnamefont {Fretel}},\
  and\ \bibinfo {author} {\bibfnamefont {P.}~\bibnamefont {Thomann}},\
  }\bibfield  {title} {\bibinfo {title} {{Bright, slow, and continuous beam of
  laser-cooled cesium atoms}},\ }\href@noop {} {\bibfield  {journal} {\bibinfo
  {journal} {Physical Review A}\ }\textbf {\bibinfo {volume} {60}},\ \bibinfo
  {pages} {4241} (\bibinfo {year} {2000})}\BibitemShut {NoStop}%
\bibitem [{\citenamefont {Skupin}\ \emph {et~al.}(2007)\citenamefont {Skupin},
  \citenamefont {Saffman},\ and\ \citenamefont {Kro}}]{Skupin2007}%
  \BibitemOpen
  \bibfield  {author} {\bibinfo {author} {\bibfnamefont {S.}~\bibnamefont
  {Skupin}}, \bibinfo {author} {\bibfnamefont {M.}~\bibnamefont {Saffman}},\
  and\ \bibinfo {author} {\bibfnamefont {W.}~\bibnamefont {Kro}},\ }\bibfield
  {title} {\bibinfo {title} {{Nonlocal Stabilization of Nonlinear Beams in a
  Self-Focusing Atomic Vapor}},\ }\href
  {https://doi.org/10.1103/PhysRevLett.98.263902} {\bibfield  {journal}
  {\bibinfo  {journal} {Physical Review Letters}\ }\textbf {\bibinfo {volume}
  {98}},\ \bibinfo {pages} {263902} (\bibinfo {year} {2007})}\BibitemShut
  {NoStop}%
\bibitem [{\citenamefont {Gibson}\ \emph {et~al.}(2020)\citenamefont {Gibson},
  \citenamefont {Baio}, \citenamefont {Robb}, \citenamefont {Ackemann},
  \citenamefont {Yao},\ and\ \citenamefont {Oppo}}]{Gibson2020}%
  \BibitemOpen
  \bibfield  {author} {\bibinfo {author} {\bibfnamefont {C.~J.}\ \bibnamefont
  {Gibson}}, \bibinfo {author} {\bibfnamefont {G.}~\bibnamefont {Baio}},
  \bibinfo {author} {\bibfnamefont {G.~R.}\ \bibnamefont {Robb}}, \bibinfo
  {author} {\bibfnamefont {T.}~\bibnamefont {Ackemann}}, \bibinfo {author}
  {\bibfnamefont {A.~M.}\ \bibnamefont {Yao}},\ and\ \bibinfo {author}
  {\bibfnamefont {G.~L.}\ \bibnamefont {Oppo}},\ }\bibfield  {title} {\bibinfo
  {title} {{Rotational dynamics of Turing patterns and cavity solitons induced
  by optical angular momentum}},\ }in\ \href
  {https://doi.org/10.1364/np.2020.nptu1d.5} {\emph {\bibinfo {booktitle}
  {Nonlinear Photonics 2020}}}\ (\bibinfo  {publisher} {OSA - The Optical
  Society},\ \bibinfo {address} {Washington, D.C.},\ \bibinfo {year}
  {2020})\BibitemShut {NoStop}%
\bibitem [{\citenamefont {Milonni}\ and\ \citenamefont
  {Boyd}(2010)}]{Milonni2010}%
  \BibitemOpen
  \bibfield  {author} {\bibinfo {author} {\bibfnamefont {P.~W.}\ \bibnamefont
  {Milonni}}\ and\ \bibinfo {author} {\bibfnamefont {R.}~\bibnamefont {Boyd}},\
  }\bibfield  {title} {\bibinfo {title} {{Momentum of Light in a Dielectric
  Medium}},\ }\href@noop {} {\bibfield  {journal} {\bibinfo  {journal}
  {Advances in Optics and Photonics}\ }\textbf {\bibinfo {volume} {2}},\
  \bibinfo {pages} {519} (\bibinfo {year} {2010})}\BibitemShut {NoStop}%
\bibitem [{\citenamefont {Fermi}\ \emph {et~al.}(1955)\citenamefont {Fermi},
  \citenamefont {Pasta},\ and\ \citenamefont {Ulam}}]{Fermi1955}%
  \BibitemOpen
  \bibfield  {author} {\bibinfo {author} {\bibfnamefont {E.}~\bibnamefont
  {Fermi}}, \bibinfo {author} {\bibfnamefont {J.}~\bibnamefont {Pasta}},\ and\
  \bibinfo {author} {\bibfnamefont {S.}~\bibnamefont {Ulam}},\ }\href@noop {}
  {\emph {\bibinfo {title} {.}}},\ \bibinfo {type} {Tech. Rep.}\ (\bibinfo
  {institution} {Los Alamos Report LA-1940},\ \bibinfo {year}
  {1955})\BibitemShut {NoStop}%
\end{thebibliography}%

\end{document}